\documentclass{aa}  

\usepackage{graphicx}
\usepackage{txfonts}
\usepackage{lipsum}
\usepackage{subcaption}         
\usepackage{lscape}             
\usepackage{placeins}           
                                
\usepackage{orcidlink}
\usepackage{amsmath}
\usepackage[dvipsnames]{xcolor}
\hypersetup{
    pdftitle={VLBI ejections},
    pdfauthor={V.~A.~Makeev},     
    colorlinks=true,  
    linkcolor=MidnightBlue,
    filecolor=magenta,      
    urlcolor=violet,
    citecolor=teal,        
    pdfborder={0 0 0}
}

\AtBeginDocument{%
  \def\figureautorefname{Fig.}%
  
}
\begin{document}

   \title{A Probabilistic Method for Estimating VLBI Jet-Component Ejection Epochs}

   \subtitle{Application to MOJAVE 15 GHz VLBA kinematics}

%
%
%

     \author{
        V.~A.~Makeev\inst{1}\corrauth{makeev.va.ph@gmail.com}{$^{\rm, }$}\thanks{Member of the International Max Planck Research School for Astronomy and Astrophysics at the Universities of Bonn and Cologne}\orcidlink{0009-0008-7830-4553}
        \and
        Y.~Y.~Kovalev\inst{1}\orcidlink{0000-0001-9303-3263}
        \and
        C.~Degli~Agosti\inst{1}{$^{\star\star}$}\orcidlink{0009-0003-6383-4950}
        \and
        D.~C.~Homan\inst{2}\orcidlink{0000-0002-4431-0890}
        }
        
        \institute{
        Max-Planck-Institut für Radioastronomie, Auf dem Hügel 69, D-53121 Bonn, Germany
        \and
        Department of Physics and Astronomy, Denison University, Granville, OH 43023, USA
        }

 
\abstract
  {Ejection epochs of parsec-scale jet components in active galactic nuclei (AGN), tracked with Very Long Baseline Interferometry (VLBI), are commonly estimated by extrapolating fitted trajectories back to the VLBI core. However, trajectory uncertainties, non-radial motion, acceleration, and the finite size and position variability of the core can make such estimates uncertain and restrict usable samples.}
  {We develop a probabilistic framework for estimating VLBI-component ejection epochs, including accelerated features, and for quantifying how consistently each backward-extrapolated trajectory is associated with an effective VLBI core region.}
  {We apply the framework to 1923 jet-component trajectories from the 15\,GHz MOJAVE kinematic sample. Ejection epochs are defined as the times of closest approach of the extrapolated trajectories to the core, with trajectory uncertainties propagated through Monte Carlo sampling. The effective core radius is treated as a random variable, whose population-level scale is calibrated from the distribution of closest-approach positions. Each component is then assigned a model-dependent probability that its closest approach lies within this effective core region.}
  {The method yields 1589 ejection epochs. The fitted effective core-region scale is $\sigma_{\rm c}=0.16\pm0.02$\,mas, corresponding to an average core radius of $0.20\pm0.03$,mas, and gives a population-level upper limit of $\lesssim0.16$,mas on the characteristic 15\,GHz core wander perpendicular to the jet. Using the same probability weighting, the full probabilistic sample has $N_{\rm eff}=624$, 2.8 times the legacy MOJAVE value of 222, while the overlapping ejection epochs remain consistent with previous estimates.}
  {The framework provides a flexible basis for population studies and cross-correlation analyses of jet structural evolution and AGN variability.}
    

   \maketitle
    \nolinenumbers
    

\section{Introduction} 

    Relativistic jets in active galactic nuclei (AGN) are highly variable systems that connect sub-parsec-scale processes near Supermassive Black Holes (SMBHs) with large-scale galaxy evolution \citep{Blandford2019}. Their emission is observed across the electromagnetic spectrum, from radio to gamma rays, and in some cases is associated with neutrino emission \citep{Abdo2010,IceCube2018}. A central difficulty in interpreting this variability is that the location of the emitting region is often uncertain.

    Very Long Baseline Interferometry (VLBI, \citealt{Thompson2017}) provides one of the strongest spatial anchors for such studies by resolving parsec-scale jet structure and tracking the apparent motion of compact jet features \citep{Kellermann2004,Lister2021,Weaver2022}. The appearance of a new VLBI component can mark an episode of activity in the jet--SMBH system, and the corresponding ejection epoch links a specific time to a physical region, usually the VLBI core, which typically is the brightest compact feature in the base of the jet \citep{Marscher1985}. Ejection epochs are therefore widely used in studies aimed at connecting jet structural evolution with radio, optical, X-ray, gamma-ray variability \citep{Savolainen2002,Marscher2008,Chatterjee2008,Jorstad2013} and even associated neutrino emission \citep{Eppel2026}.

    However, most ejection-epoch estimates rely on deterministic back-extrapolation of fitted feature trajectories to the VLBI core. Depending on the adopted kinematic description, these include linear trajectories \citep{Lister2019}, piecewise-linear trajectories with extrapolation of the innermost segment \citep{Weaver2022}, and polynomial trajectory fits \citep{Jorstad2017}. Such estimates can be sensitive to trajectory uncertainties, non-radial motion, significant acceleration, and the finite effective size of the VLBI core. Another significant effect that is usually neglected is the wander of the VLBI core itself, alternatively referred to as the core shuttle effect \citep{Lisakov2017, Plavin2019}. These effects are especially important in population studies, where uncertain or marginal ejection events should not necessarily be discarded, but should also not be treated as equally reliable.
    
    In this work, we present a probabilistic framework for estimating ejection epochs of VLBI jet components from fitted trajectory models. The method defines the ejection epoch through the closest-approach position of the extrapolated trajectory to the VLBI core. It is applicable to features with significant acceleration and, additionally, assigns each event a model-dependent ejection probability based on the effective core region, accounting for the finite VLBI core size and its motion. We apply the framework to the 15~GHz MOJAVE (Monitoring Of Jets in Active galactic nuclei with VLBA Experiments) kinematic sample \citep{Lister2018,Lister2021} as a large homogeneous test case, compare the results with previous conservative estimates, and discuss how the resulting probabilities can be used in statistical studies of AGN variability.
    
    The paper is organized as follows. In \autoref{s:framework}, we introduce the statistical framework for estimating ejection epochs, including the closest-approach definition, Monte Carlo uncertainty propagation, and the ejection probability. In \autoref{s:application}, we apply the method to the MOJAVE 15~GHz VLBA kinematic data set, calibrate the effective core-region scale, and present the resulting ejection epochs and probabilities. In \autoref{s:discussion}, we compare the results with the legacy MOJAVE estimates and discuss the interpretation, advantages, and limitations of the method. We present the main conclusions and outline possible applications and extensions in \autoref{s:summary}.

\section{Statistical Framework for Ejection Epoch Estimation}
\label{s:framework}
    
    \subsection{Definition and Estimation of Ejection Epochs}

    We describe the motion of a jet feature in the image plane by the position vector $\mathbf{r}(t)$, while $\mathbf{r}_{\rm core}$ denotes the model-fitted VLBI-core position. The instantaneous separation between the feature and the core is
    \begin{equation}
    d(t) = \left\lVert \mathbf{r}(t) - \mathbf{r}_{\rm core} \right\rVert.
    \end{equation}
    For each fitted trajectory, we operationally define the ejection epoch as the kinematic closest-approach epoch,
    \begin{equation}
    t_{\rm ej} = \operatorname*{arg\,min}_{t < t_{\rm first}} d(t),
    \end{equation}
    where the minimization is restricted to epochs preceding the first observing epoch, $t_{\rm first}$. If no minimum exists within this interval, no ejection epoch is assigned to the trajectory. The corresponding closest-approach position and distance are
    \begin{equation}
    \mathbf{r}_{\rm ca} = \mathbf{r}(t_{\rm ej}), \qquad d_{\rm ca} = \left\lVert \mathbf{r}_{\rm ca} - \mathbf{r}_{\rm core} \right\rVert.
    \end{equation}
    The associated ejection probability introduced in \autoref{s:probability} quantifies the probability that this closest approach lies within the effective VLBI-core region (i.e. the region, in which VLBI-core might be fully localized, given its potential position and size variability; for details see \autoref{s:discussion_interp}). The correspondence between the kinematic closest-approach epoch and the physical launch time assumes that any unresolved initial acceleration phase is short compared with the extrapolation interval. In particular, an unresolved phase of monotonic outward acceleration would generally cause the extrapolated epoch to occur later than the physical launch time.
    
    Following \citet{Lister2019,Lister2021}, we assume that the motion of each feature is described either by a constant-velocity or a constant-acceleration model. In the image plane, the trajectory is written as
    \begin{equation}
    \label{eq:traj_vector}
    \mathbf{r}(t) = \mathbf{r}_0 + \mathbf{v}\,(t-t_{\rm mid}) + \frac{1}{2}\,\mathbf{a}\,(t-t_{\rm mid})^2,
    \end{equation}
    where $\mathbf{r}_0$ is the position of the feature at the middle observing epoch $t_{\rm mid}$, $\mathbf{v}=(v_x,v_y)$ is the velocity vector, and $\mathbf{a}=(a_x,a_y)$ is the acceleration vector. All positions are measured in the image plane relative to the fitted VLBI-core position, such that $\mathbf{r}_{\rm core}=\mathbf{0}$.
    
    For features without significant acceleration, $\mathbf{a}=\mathbf{0}$, the trajectory is linear and $t_{\rm ej}$ can be obtained analytically by minimizing $\lVert\mathbf{r}(t)\rVert^2$. At the point of closest approach, the position vector is perpendicular to the velocity vector, such that $\mathbf{r}(t_{\rm ej})\cdot\mathbf{v}=0$. This yields
    \begin{equation} \label{eq:ca_vect}
    \left\{
    \begin{aligned}
    t_{\rm ej} &= t_{\rm mid} - \frac{\mathbf{r}_0\cdot\mathbf{v}}{\lVert\mathbf{v}\rVert^2}, \\
    \mathbf{r}_{\rm ca} &= \mathbf{r}_0 + \mathbf{v}\,(t_{\rm ej}-t_{\rm mid}), \\
    d_{\rm ca} &= \lVert\mathbf{r}_{\rm ca}\rVert.
    \end{aligned}
    \right.
    \end{equation}
    
    For features with significant acceleration, no closed-form expression for $t_{\rm ej}$ is used, and the minimum of $\lVert\mathbf{r}(t)\rVert$ is determined numerically. The numerical search is initialized at the constant-velocity closest-approach epoch given by \autoref{eq:ca_vect}, and therefore preferentially converges to the stationary point continuously connected to the linear back-extrapolation solution. For accelerated trajectories, multiple local minima may in principle exist, but minima occurring after the first observing epoch are outside the adopted ejection-epoch domain. The candidate closest-approach epochs are obtained by solving a cubic equation
    \begin{equation}
    \frac{d}{dt}\lVert\mathbf{r}(t)\rVert^2 = 0,
    \end{equation}
    after which the corresponding closest-approach position and distance, $r_{\rm ca}$ and $d_{\rm ca}$, are evaluated from the trajectory.
    
    To estimate uncertainties in $t_{\rm ej}$, $\mathbf{r}_{\rm ca}$, and $d_{\rm ca}$, we employ a Monte Carlo approach. The fitted trajectory parameters, $\mathbf{r}_0$, $\mathbf{v}$, and $\mathbf{a}$ where applicable, are assumed to be normally distributed, with standard deviations given by their reported uncertainties. For each realization, a set of trajectory parameters is drawn from these distributions, and the corresponding values of $t_{\rm ej}$, $\mathbf{r}_{\rm ca}$, and $d_{\rm ca}$ are computed analytically or numerically, depending on the adopted motion model. Repeating this procedure 5000 times yields empirical distributions for all three quantities. We characterise these distributions by their median values and percentile-based confidence intervals. In particular, we report the 16th--84th percentiles, corresponding approximately to a $1\sigma$ interval for a Gaussian distribution, and, where relevant, the 2nd--98th percentiles. Additional summary statistics, including the mean and standard deviation, are also computed.

    \subsection{Ejection Probability and Core Region Modelling}
    \label{s:probability}

    We represent the effective VLBI-core region in the image plane as a circular disk centred on the fitted core position, with radius $R_{\rm c}\geq 0$. The event associated with an ejection is defined as the closest-approach distance lying within this region. Because both the trajectory and the effective extent of the core region are uncertain, we treat the closest-approach distance, $D_{\rm ca}\geq 0$, and the core radius, $R_{\rm c}\geq 0$, as random variables. Their probability density functions are denoted by $\rho_{\rm ca}(d)$ and $\psi_{\rm c}(r_{\rm c})$, respectively, where $d$ and $r_{\rm c}$ are non-negative scalar distances.
    
    Conditional on the adopted core-region scale parameter $\sigma_{\rm c}$, the trajectory-derived closest-approach distance and the effective core radius are assumed to be statistically independent,
    \begin{equation}
    p(D_{\rm ca},R_{\rm c}\mid\mathcal{D},\sigma_{\rm c})
    =
    p(D_{\rm ca}\mid\mathcal{D})\,p(R_{\rm c}\mid\sigma_{\rm c}),
    \end{equation}
    where $\mathcal{D}$ denotes the fitted trajectory parameters and their uncertainties. We define the ejection probability as the conditional probability that the closest-approach distance is smaller than the effective core radius,
    \begin{equation}
    P_{\rm ej}
    =
    \Pr(D_{\rm ca}\leq R_{\rm c}\mid\mathcal{D},\sigma_{\rm c}).
    \end{equation}
    Under the conditional-independence assumption, this probability can be written as
    \begin{equation}
    \begin{split}
    P_{\rm ej}
    &=
    \int_0^\infty \rho_{\rm ca}(d)\,
    \Pr(R_{\rm c}\geq d\mid\sigma_{\rm c})\,{\rm d}d \\
    &=
    \int_0^\infty \rho_{\rm ca}(d)
    \left[
    \int_d^\infty \psi_{\rm c}(r_{\rm c}\mid\sigma_{\rm c})\,{\rm d}r_{\rm c}
    \right]{\rm d}d.
    \end{split}
    \label{eq:pej_general}
    \end{equation}
    
    We model the effective core radius using a Rayleigh distribution. This distribution naturally arises when a non-negative radial quantity is interpreted as the magnitude of an isotropic two-dimensional Gaussian random vector. It therefore provides a simple isotropic model for the uncertain effective extent of the core region. The corresponding probability density function is
    \begin{equation}
    \psi_{\rm c}(r_{\rm c}\mid\sigma_{\rm c})
    =
    \frac{r_{\rm c}}{\sigma_{\rm c}^2}
    \exp\left(-\frac{r_{\rm c}^2}{2\sigma_{\rm c}^2}\right),
    \qquad r_{\rm c}\geq 0,
    \end{equation}
    where $\sigma_{\rm c}$ is the core-region scale parameter. The corresponding survival function is
    \begin{equation}\label{eq:core_probability}
    \Pr(R_{\rm c}\geq d\mid\sigma_{\rm c})
    =
    \exp\left(-\frac{d^2}{2\sigma_{\rm c}^2}\right).
    \end{equation}
    
    The exact functional form of $\rho_{\rm ca}(d)$ is not known analytically, but it can be sampled through the Monte Carlo propagation of the trajectory uncertainties. Let $\{d_{{\rm ca},i}\mid i=1,\ldots,N_{\rm MC}\}$ denote the resulting closest-approach distances. The ejection probability can then be estimated as
    \begin{equation}
    \label{eq:p_ej}
    \begin{split}
    P_{\rm ej}
    &=
    \int_0^\infty \rho_{\rm ca}(d)
    \exp\left(-\frac{d^2}{2\sigma_{\rm c}^2}\right)\,{\rm d}d \\
    &\approx
    \frac{1}{N_{\rm MC}}
    \sum_{i=1}^{N_{\rm MC}}
    \exp\left(-\frac{d_{{\rm ca},i}^2}{2\sigma_{\rm c}^2}\right).
    \end{split}
    \end{equation}

    \begin{figure*}[t]
    \centering
    \includegraphics[width=\textwidth]{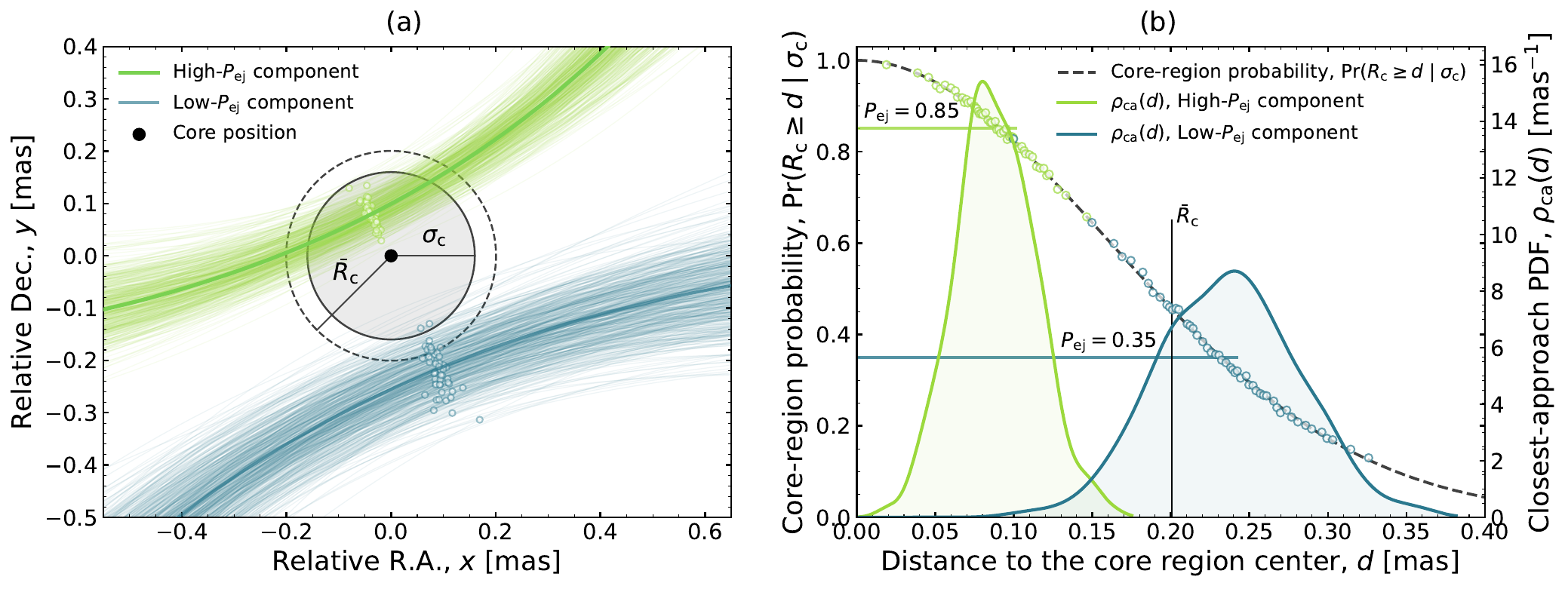}
    \caption{
        Illustration of the probabilistic ejection-epoch framework for two representative components with high and low $P_{\rm ej}$ (see \autoref{s:probability}). 
        (a) Monte Carlo realizations of mock accelerated trajectories extrapolated toward the VLBI core. The coloured lines show individual trajectory realizations for the high- and low-$P_{\rm ej}$ components, respectively. The shaded region represents the effective core region with $\sigma_{\rm c}=0.16$~mas and mean radius $\bar{R}_{\rm c}$, as calibrated in \autoref{s:core_est}. Markers show the corresponding closest-approach positions, with distances $d_{\rm ca}$ from the core centre. 
        (b) Conversion of the sampled $d_{\rm ca}$ values into $P_{\rm ej}$. The dashed curve shows the core-region probability $\Pr(R_{\rm c}\geq d\mid\sigma_{\rm c})$, the black vertical line marks $\bar{R}_{\rm c}$, and the coloured curves show the closest-approach distributions $\rho_{\rm ca}(d)$. Open circles illustrate the mapping of individual $d_{\rm ca}$ values onto core-region probabilities; the resulting $P_{\rm ej}=0.85$ and $0.35$ are shown by the horizontal coloured lines.
    }
    \label{fig:ejection_framework}
    \end{figure*}
    
    Thus, for each fitted trajectory, $P_{\rm ej}$ gives the model-dependent probability that its closest approach lies within the effective core region, conditional on the trajectory uncertainties and the adopted core-region scale parameter $\sigma$. A detailed illustration showing how Monte Carlo sampled trajectories are processed and converted into the ejection probabilities is presented in \autoref{fig:ejection_framework}.

    Our approach can also be viewed as a direct generalization of the case of a fixed core radius. If the core radius is assumed to have a fixed value $\bar{R}_{\rm c}$, its probability density is simply a delta function,
    \begin{equation}
        \psi_{\rm const}(r_{\rm c}\mid \bar{R}_{\rm c})
        =
        \delta(r_{\rm c}-\bar{R}_{\rm c}).
    \end{equation}
    In this case, \autoref{eq:pej_general} reduces to
    \begin{equation}
    \label{eq:p_ej_const}
    \begin{split}
    P_{\rm ej,\,const}
    &=
    \int_0^\infty \rho_{\rm ca}(d)
    1_{[0,\bar{R}_{\rm c}]}(d)\,{\rm d}d \\
    &\approx
    \frac{1}{N_{\rm MC}}
    \sum_{i=1}^{N_{\rm MC}}
    1_{[0,\bar{R}_{\rm c}]}(d_{{\rm ca},i}),
    \end{split}
    \end{equation}
    where $1_{[0,\bar{R}_{\rm c}]}(d)$ is equal to 1 for $d\leq \bar{R}_{\rm c}$ and 0 otherwise. Thus, in the fixed-radius case, $P_{\rm ej}$ is simply the fraction of Monte Carlo trajectory realizations whose closest approach lies within the adopted core radius. In the limit of a perfectly constrained trajectory originating within the core region, the ejection probability is equal to unity.
    
    Comparing \autoref{eq:p_ej} and \autoref{eq:p_ej_const} also provides a useful connection between the core-region scale parameter $\sigma_{\rm c}$ and the radius $\bar{R}_{\rm c}$ in the fixed-core case. Requiring the two weighting functions to have the same integrated weight gives
    \begin{equation}
        \bar{R}_{\rm c}
        =
        \sqrt{\frac{\pi}{2}}\,\sigma_{\rm c}.
        \label{eq:rc_sigma_relation}
    \end{equation}
    Notably, the right-hand side is exactly the mean core radius, $\langle R_{\rm c}\rangle$, for the Rayleigh distribution adopted in our model. This relation therefore provides a direct correspondence between the probabilistic core scale and a fixed core radius.

\section{Application to MOJAVE data}
\label{s:application}
    
    \subsection{Kinematics of Jet Components from 15 GHz VLBI Observations}\label{ss:mojave}
    
       We test and validate the method using the MOJAVE 15\,GHz VLBA kinematic data set \citep{Lister2021}. This sample provides a large and homogeneous collection of parsec-scale jet-component trajectories, making it well suited for testing a population-level ejection-epoch estimator. The MOJAVE archive contains approximately ten thousand 15\,GHz VLBA observations of more than 500 AGN, and the kinematic analysis of \citet{Lister2021} covers 447 sources observed between 1994 August 31 and 2019 August 4.

        For each source, the sky brightness distribution is modelled directly in the $(u,v)$ plane using the \texttt{modelfit} task in \texttt{DIFMAP} \citep{Shepherd1997,Lister2021}. Individual jet features are cross-identified across epochs based on their evolution in position, flux density, and brightness temperature. The bright VLBI core is assumed to be stationary and is used as the reference point for the measured component positions.
        
        The final decision on cross-identification is made by several members of the MOJAVE group. Additionally, each feature is discussed and, based on the model-fit quality, its flux level, and general dynamics, may be considered robust or non-robust. Only robust cross-identified trajectories are then fitted using either a constant-velocity or a constant-acceleration model, as described generally by \autoref{eq:traj_vector}. A feature is considered to have significant acceleration only if it is observed at more than ten epochs and the acceleration estimate is significantly larger than its uncertainty, $a > 3\sigma_a$. In this case, the motion is described by the full quadratic form in \autoref{eq:traj_vector}. Otherwise, the acceleration is set to zero and the motion is modelled using a non-accelerating two-dimensional vector fit \citep{Lister2021}. Also, for a subset of features, their analysis includes ejection epochs, described in detail in \autoref{s:comparison_legacy}.

        \subsection{Estimation of the typical parameters of the core}
        \label{s:core_est}
        Although the angular size and position of the VLBI core can be estimated from model fitting \citep[e.g.,][]{Homan2021,Lister2021}, they do not necessarily reflect the region from which the jet-component trajectories emerge. It was previously reported for the quasar 3C~273, based on an analysis of individual jet-component trajectories relative to the fitted VLBI-core position in 43~GHz observations, that the core may actually wander \citep{Lisakov2017}. Additionally, significant core-shift variability was observed between 8~GHz and 2.4~GHz VLBI observations for a sample of AGNs, directly implying variability of the VLBI-core position at these frequencies \citep{Plavin2019}. Given that jet-component positions relative to the VLBI core are sensitive to such motions, we can use them to constrain the effective region within which the VLBI core may actually be located. In the approximation discussed in \autoref{s:probability}, we therefore estimate a representative population-level core-region scale parameter, $\sigma_{\rm c}$, using all 1284 non-accelerated component trajectories. For each trajectory from \citet{Lister2021}, we compute the position of closest approach to the VLBI core, $\mathbf{r}_{\rm ca}=(x_{\rm ca},y_{\rm ca})$, where $x$ and $y$ correspond to the relative right ascension ($\Delta \rm RA$) and declination ($\Delta \rm Dec$) offsets, respectively.

\begin{figure}
            \centering
            \includegraphics[width=0.95\linewidth]{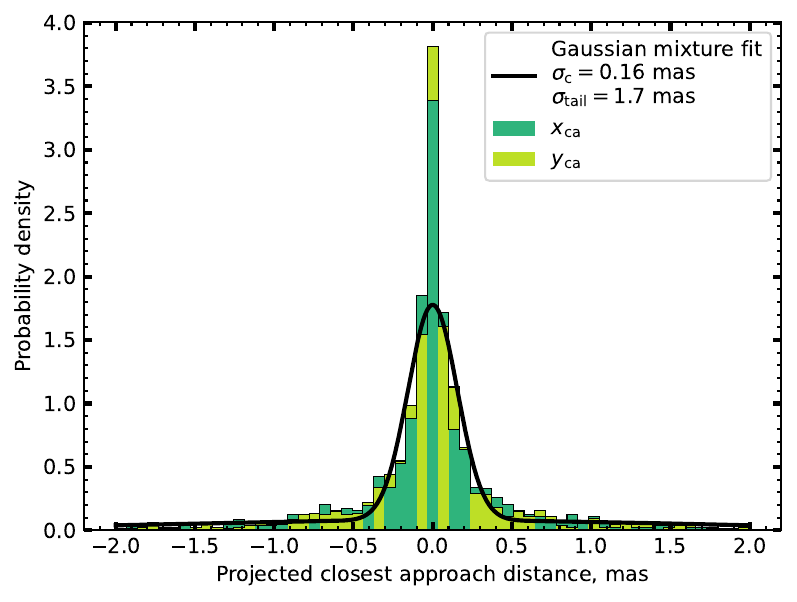}
            \caption{Distribution of closest-approach positions $\mathbf{r}_{\rm ca}=(x_{\rm ca},y_{\rm ca}) = (\Delta \rm RA, \Delta \rm Dec)$ for all 1284 non-accelerated component trajectories. The best-fit values of $\sigma_{\rm c}$ and $\sigma_{\rm tail}$ of the Gaussian mixture are shown in the legend. For clarity, only the central region of the distribution, $|x_{\rm ca}|, |y_{\rm ca}| \leq 2$~mas, is shown.}
            \label{fig:r_min_fit}
        \end{figure}

        The resulting distribution (Fig.~\ref{fig:r_min_fit}) is centred at an offset of $0$~mas, as expected, since all feature coordinates are estimated relative to the VLBI core position at observed epochs. It also exhibits extended tails, which are primarily caused by large uncertainties in the trajectory parameters of some components. To obtain a robust estimate of the central distribution, we first remove outliers by selecting the inner 99\% of the $|r_{\rm ca}|$ distribution. To account for the remaining extended tails, we model the truncated distribution with a circular, zero-centred Gaussian mixture, i.e., a weighted sum of two Gaussian components. Since different sources can have significantly different numbers of jet components, we additionally weight the contribution of each source to ensure that all sources contribute equally to the final estimates. The uncertainties are estimated by bootstrapping the sources over 2000 iterations.

        For the best-fit model, the narrow component has a standard deviation of $\sigma_{\rm c} = 0.16 \pm 0.02$~mas and a weight of $0.66 \pm 0.03$. The broad component, which accounts for the extended tails, has $\sigma_{\rm tail} = 1.7 \pm 0.2$~mas. The probability density of the best-fit distribution is also shown in Fig.~\ref{fig:r_min_fit}. Additionally, based on \autoref{eq:rc_sigma_relation}, we estimate the average core radius as
        $
        \bar{R}_{\rm c}
        =
        0.20\pm0.03~\mathrm{mas}.
        $
        Minor variation of the distribution truncation threshold changes the final result only modestly.
        
        In the following analysis, we treat $\sigma_{\rm c}$ as a population-level calibration parameter rather than as a quantity inferred separately for each trajectory. It is derived from the global distribution of closest-approach positions and therefore captures the average effective scale of the core region under the adopted trajectory models. For a given value of $\sigma_{\rm c}$, the corresponding $P_{\rm ej}$ therefore measures, within this population-level core model, the probability that an individual feature trajectory is consistent with originating from the effective core region.

        The distribution of $\mathbf{r}_{\rm ca}$ and the best-fit Gaussian contours are shown in \autoref{fig:r_min_fit}. The substantial spread of the observed distribution indicates that adopting a fixed core size is not adequate, and supports the use of a probabilistic (random-variable) description of the core radius. Notably, the derived value of $\bar{R}_{\rm c}$ is consistent with the core size adopted in \citet{Lister2021}, providing a useful consistency check for the method. With this calibration fixed, the resulting $P_{\rm ej}$ values are conditional ejection probabilities under the adopted MOJAVE trajectory models and the population-level effective core-region model.

        \subsection{Ejection epochs and their probabilities}

        Using the estimated typical core scale parameter $\sigma_{\rm c}$, we analyse all 1923 trajectory fits from \citet{Lister2021}. Of these, 1589 have a closest-approach minimum preceding the first observing epoch, $t_{\rm ej}<t_{\rm first}$, and are therefore assigned an ejection epoch and included in the resulting catalogue presented in \autoref{tab:ej_catalog_simple}. In that work, accelerated-motion models were adopted for 304 trajectories with statistically significant acceleration, while vector-motion models were used for the remaining 1285.

        \begin{figure}
        \centering
        \includegraphics[width=0.99\linewidth]{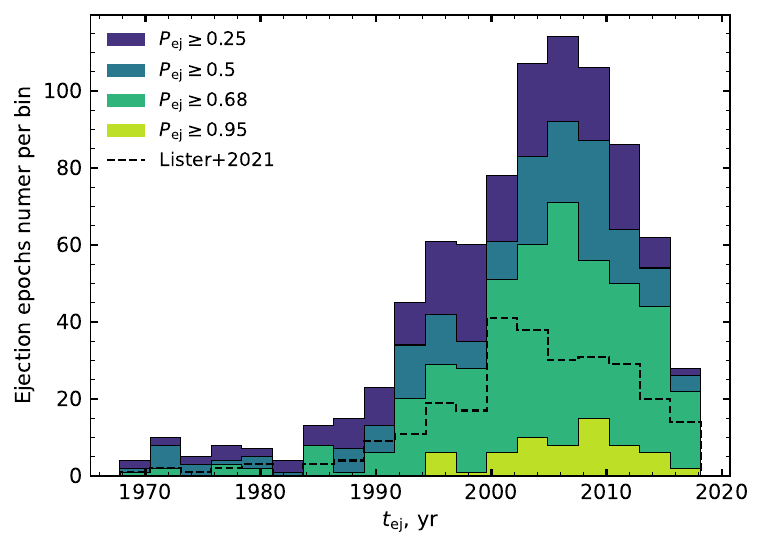}
        \caption{Distribution of ejection epochs. The colour indicates different ejection probability values, $P_{\rm ej}$. The dashed line shows the distribution of ejection epochs reported in \citet{Lister2021}. For clarity, values with $t_{\rm ej}<1965$ are omitted from the plot; this affects only 17 estimates and has a negligible statistical contribution.}
        \label{fig:t_ej_dist}
        \end{figure}
        
        The derived ejection epochs span from the mid-20th century to the end of 2019, with the majority concentrated between 1990 and 2019. The distribution of ejection epochs for different minimum probability thresholds $P_{\rm ej}$ is shown in \autoref{fig:t_ej_dist}. Each ejection epoch should be considered together with its associated probability. For simplicity, one may adopt a reference fiducial threshold $P_{\rm ej} \geq 0.68$, motivated by the 68\% enclosed-probability convention for a Gaussian distribution. In this case, the median ejection epoch is
        $
        t_{{\rm ej,\,med}} = 2005.8~\mathrm{yr},
        $
        with 90\% of ejection epochs being between $1991.7~\mathrm{yr}$ and  $2015.5~\mathrm{yr}$.

        \begin{figure}
        \centering
        \includegraphics[width=0.99\linewidth]{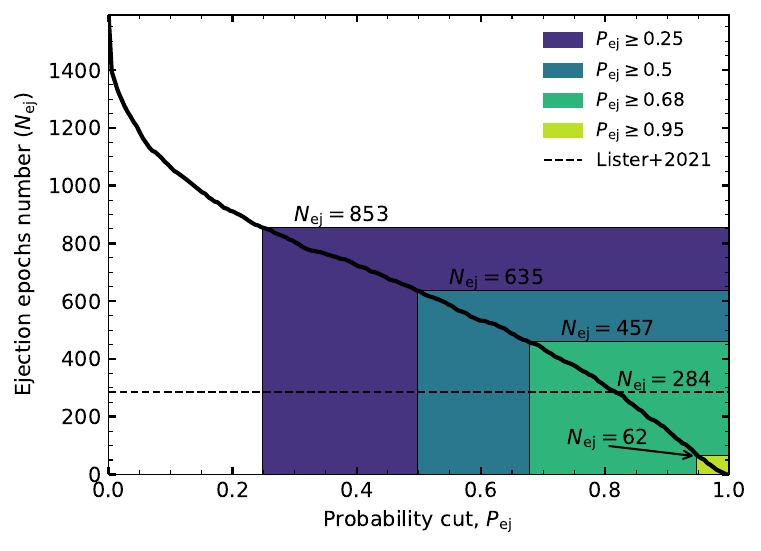}
        \caption{Cumulative distribution of the number of ejection epochs, $N_{\rm ej}$, as a function of the probability cut, $P_{\rm ej}$. The black curve shows the cumulative distribution. Coloured shaded regions indicate the ranges of $P_{\rm ej}$ thresholds and the corresponding numbers of selected ejection epochs, $N_{\rm ej}$. The dashed horizontal line marks the total number of ejection epochs reported in \citet{Lister2021}. Values of $N_{\rm ej}$ for representative cuts are annotated on the plot. }
        \label{fig:t_ej_p_ej}
        \end{figure}

        As seen in \autoref{fig:t_ej_dist}, the inferred distribution depends strongly on the adopted probability threshold. To quantify this effect, we examine the dependence of the total number of ejection epochs, $N_{\rm ej}$, on the minimum probability cut, as shown in \autoref{fig:t_ej_p_ej}. Out of the full sample of 1589 ejection epochs, only about half have $P_{\rm ej} > 0.25$, and approximately 40\% have $P_{\rm ej} > 0.5$. This indicates that a large fraction of the fitted trajectories do not extrapolate sufficiently close to the effective core region under the adopted kinematic models. This behaviour is consistent with the prevalence of non-radial and complex motions reported in previous VLBI studies \citep{Lister2019, Lister2021}.
        For the fiducial threshold ($P_{\rm ej} \geq 0.68$), the number of selected ejection epochs is $N_{\rm ej} = 457$, while for the more stringent criterion, $P_{\rm ej} \geq 0.95$, only 62 ejection epochs remain. 
        
        Since $P_{\rm ej}$ is defined individually for each trajectory, we additionally characterise a sample by its probability-weighted effective number of ejections,
        \begin{equation}
            N_{\rm eff}=\sum_i P_{{\rm ej},i}.
        \end{equation}
        For all 1589 estimated ejection epochs, $N_{\rm eff}=624$. For the $P_{\rm ej}\ge0.68$ subset, the 457 selected epochs have a mean probability of $0.85$, corresponding to $N_{\rm eff}=387$.

        \begin{figure}
            \centering
            \includegraphics[width=0.99\linewidth]{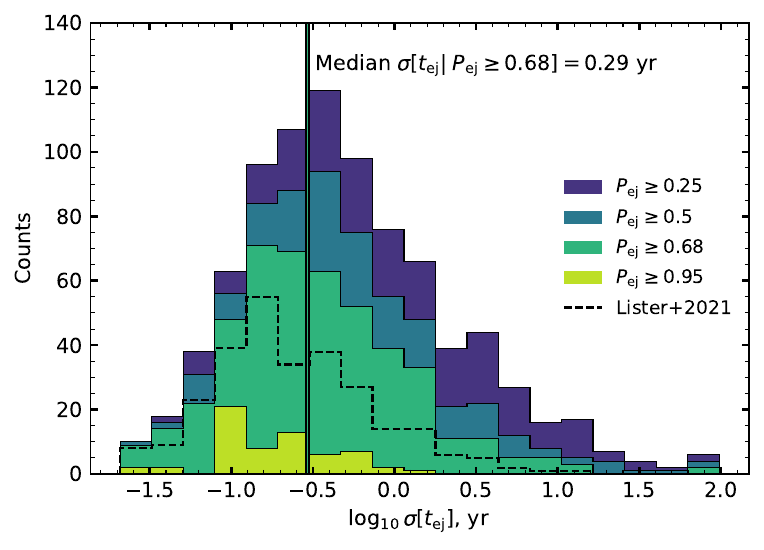}
            \caption{Distribution of uncertainties in ejection epochs. For simplicity, we adopt symmetrized uncertainties defined as half of the difference between the 84th and 16th percentiles of each $t_{\rm ej}$ posterior distribution, which is equivalent to the $1\sigma$ value for a Gaussian distribution. Different colours correspond to different cuts on the ejection probability, $P_{\rm ej}$. The dashed histogram shows the distribution of uncertainties reported in \citet{Lister2021}. The green vertical line (with a black edge) marks the median uncertainty for features with $P_{\rm ej} \geq 0.68$.}
            \label{fig:errors}
        \end{figure}
        
        The distribution of uncertainties in the ejection epochs is shown in \autoref{fig:errors}. For clarity, we use symmetrized uncertainties defined as half of the difference between the 84th and 16th percentiles of the posterior distribution of $t_{\rm ej}$. In the Gaussian case, this corresponds to the standard $1\sigma$ uncertainty. 
        A comparison with the full (asymmetric) uncertainties from \autoref{tab:ej_catalog_simple} shows that the lower errors are, on average, about 20\% larger; i.e., the $t_{\rm ej}$ posteriors are left-skewed. Nevertheless, the interval between the 16th and 84th percentiles closely matches the symmetric estimate $t_{\rm ej} \pm \sigma$, with deviations below one month for more than 90\% of the sample.
        
        Most ejection epochs, regardless of probability, are constrained within $\sim$1 year. However, the spread of the uncertainty distribution increases significantly toward lower $P_{\rm ej}$ values. This trend indicates that the ejection probability serves as a meaningful additional proxy for the reliability of the estimated ejection epoch. For the subset with $P_{\rm ej} \geq 0.68$, the median uncertainty together with upper and lower boundaries from 84th and 16th percentiles is
        $
        \sigma_\mathrm{med}[{t_{\rm ej}} |~P_{\rm ej} \geq 0.68] = 0.29_{-0.19}^{+0.92}~\mathrm{yr}.
        $
        Larger uncertainties are typically associated with poorly constrained trajectory parameters, for example in the case of slowly moving components or when extrapolation to the ejection epoch involves long time intervals. We additionally compared the ejection-epoch uncertainties for accelerated and non-accelerated components with $P_{\rm ej}>0.68$, using only components with more than 10 observing epochs to make the comparison uniform. The median uncertainties are $0.20$~yr and $0.24$~yr, respectively. The slightly smaller errors for accelerated fits are expected, since the more flexible trajectory model can follow the component motion more closely.

\section{Discussion}
\label{s:discussion}

            \renewcommand{\arraystretch}{1.15}
            \setlength{\tabcolsep}{10pt}
            
            \begin{table*}
            \caption{The ejection epoch catalogue, produced on the basis of kinematics results from \citet{Lister2021}.}
            \label{tab:ej_catalog_simple}
            \centering
            \begin{tabular}{l c c c c c c c}
            \hline\hline
            Source &
            Feature &
            $N_{\rm obs}$ &
            $t_{\rm mid}$, yr &
            $P_{\rm ej}$ &
            $t_{\rm ej}$, yr &
            $x_{\rm ca}$, $\mu$as &
            $y_{\rm ca}$, $\mu$as \\
            (1) & (2) & (3) & (4) & (5) & (6) & (7) & (8) \\
            \hline
            0003+380 & 1        & 8  & 2008.81 & $0.05$  & $1986.1^{+5.4}_{-8.4}$   & $-66^{+684}_{-205}$   & $-1703^{+1272}_{-1046}$ \\
                     & 2        & 6  & 2007.71 & 0.41                  & $2002.08^{+0.37}_{-0.43}$   & $114^{+78}_{-65}$     & $198^{+91}_{-97}$ \\ \\
                                        
            0003-066 & 3        & 9  & 1999.33 & $0.09$   & $1988.9^{+1.5}_{-1.9}$   & $137^{+17}_{-39}$     & $564^{+248}_{-269}$ \\
                     & $4^{\rm a}$  & 26 & 2004.83 & $0$                   & $1985.99^{+0.67}_{-0.73}$   & $-765^{+223}_{-243}$  & $2995^{+239}_{-267}$ \\
                     & $5^{\rm a}$  & 14 & 2004.37 & 0.38                  & $1999.18^{+0.42}_{-0.52}$   & $-112^{+21}_{-15}$    & $200^{+77}_{-73}$ \\
                     & $8^{\rm a}$  & 12 & 2009.93 & 0.62                  & $2004.63^{+0.24}_{-0.28}$   & $-37^{+78}_{-109}$    & $-68^{+161}_{-146}$ \\
                     & 9        & 10 & 2009.24 & 0.45                  & $2003.24^{+0.48}_{-0.52}$   & $-93^{+65}_{-81}$     & $-196^{+125}_{-111}$ \\
                     & 13       & 6  & 2011.33 & 0.89                  & $2002.98^{+0.99}_{-1.25}$   & $-9^{+78}_{-76}$      & $3^{+38}_{-18}$ \\
                     & 14       & 7  & 2011.08 & $0.08$   & $1999.6^{+2.1}_{-2.8}$   & $-381^{+229}_{-281}$  & $-459^{+176}_{-73}$ \\
            \hline
            \end{tabular}
            
            \tablefoot{
                Columns are as follows: (1) B1950 source name; (2) feature identification number;
                (3) number of observing epochs used to fit the feature trajectory;
                (4) middle observing epoch, defined as $(t_{\rm max}+t_{\rm min})/2$;
                (5) ejection probability $P_{\rm ej}$, defined in \autoref{s:probability};
                (6) median estimated ejection epoch $t_{\rm ej}$, defined as the epoch of closest approach of the extrapolated trajectory to the VLBI core (see \autoref{s:framework});
                (7) median relative right ascension offset $x_{\rm ca}=\Delta{\rm RA}$ of the trajectory at the point of closest approach;
                (8) median relative declination offset $y_{\rm ca}=\Delta{\rm Dec}$ of the trajectory at the point of closest approach.
                For $t_{\rm ej}$, $x_{\rm ca}$, and $y_{\rm ca}$, uncertainties correspond to the 16th and 84th percentiles of their Monte Carlo distributions.
                $^{a}$ Feature shows significant acceleration.
                The full catalogue in electronic form is available at the CDS. We show here the first two sources as a guide to its form and content.
                }
            \end{table*}

    \subsection{Comparison with previous ejection-epoch estimates}\label{s:comparison_legacy}
        Different approaches have previously been used to account for non-linear component motion when estimating ejection epochs. \citet{Jorstad2017} extrapolated polynomial trajectories toward the core; for higher-order trajectories, the ten epochs closest to the core were refitted with a polynomial of order $l\leq2$ for this purpose. \citet{Weaver2022} instead used piecewise-linear trajectories and extrapolated the innermost linear segment. In both cases, the core-crossing epochs were determined separately along the two image-plane coordinates and combined into a single uncertainty-weighted ejection epoch. 
        
        For a direct quantitative comparison, we use the legacy ejection epochs reported in \citet{Lister2021}, which were obtained using a more conservative selection. To qualify for an ejection epoch estimate, a feature must be labelled as robust (see \autoref{ss:mojave} for explanation), show significant motion ($v \ge 3\sigma_v$), no significant acceleration, and a velocity vector directed within $15^\circ$ of the outward radial direction \citep{Lister2021}. Additionally, its extrapolated position at the ejection epoch must be within 0.2~mas of the core \citep{Lister2021}. The relative RA and Dec motions are extrapolated independently to zero offsets, yielding two ejection-time estimates. These are then combined as an inverse-variance weighted average.

        {\renewcommand{\figureautorefname}{Figs.}%
        \autoref{fig:t_ej_dist} and~\ref{fig:t_ej_p_ej}} demonstrate the increase in statistical power relative to the ejection epochs reported by \citet{Lister2021}. Their 284 ejection epochs have a mean probability of $0.78$ in our framework, corresponding to $N_{\rm eff}=222$, showing that their conservative selection preferentially retains high-probability events, but at the cost of a substantially smaller sample. In comparison, the $P_{\rm ej}\ge0.68$ selection contains 457 epochs with a mean probability of $0.85$ and $N_{\rm eff}=387$, representing a 60\% increase in the raw sample size and a $74$\% increase in the probability-weighted effective number. Using all estimated epochs gives $N_{\rm eff}=624$, approximately $2.8$ times that of the legacy sample.
        
        \begin{figure}
            \centering
            \includegraphics[width=0.99\linewidth]{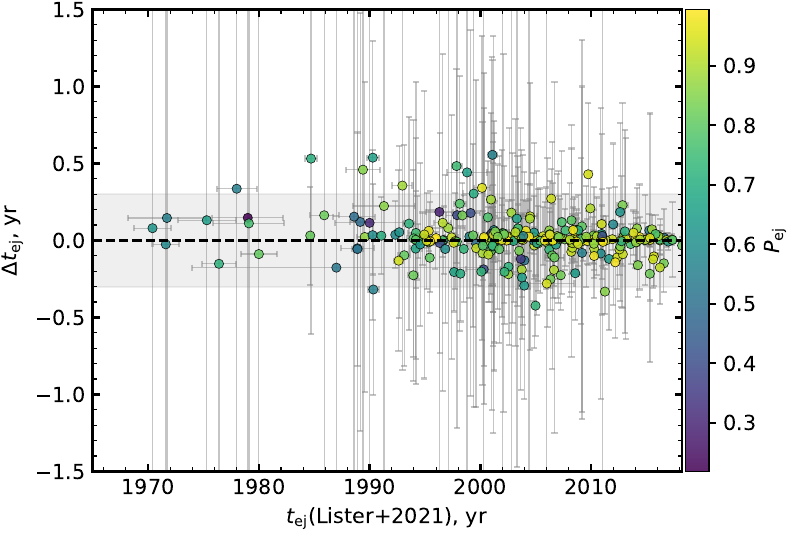}
            \caption{Comparison of ejection epochs estimated in this work with those reported in \citet{Lister2021}. On $y$-axis the difference  $\Delta t_{\rm ej} = t_{\rm ej} - t_{\rm ej,\,Lister}$ is shown, with combined uncertainties $\sigma_{\rm comb} = \sqrt{\sigma_{\rm ej,\,new}^2 + \sigma_{\rm ej,\,Lister}^2}$. The grey shaded region indicates the central 95\% of absolute differences, corresponding to $\pm 0.3$~yr. The colour scale represents the ejection probability, $P_{\rm ej}$, derived in this work. For clarity, values with $t_{\rm ej}<1965$ are omitted from the plot; this affects only 3 estimates and has a negligible statistical contribution.}
            \label{fig:errors_cmp}
        \end{figure}
        
        The uncertainties derived in this work are in good agreement with those reported in \citet{Lister2021}, with a median relative difference of only 0.2\% and about 5\% differing by more than 15\% (\autoref{fig:errors}). Within the overlap of the two datasets, the estimated ejection epochs are likewise consistent within their uncertainties, with a median absolute difference of 0.04~yr and 95\% of the differences below 0.3~yr (\autoref{fig:errors_cmp}).
        
        Overall, these results demonstrate strong consistency between the two approaches, while the probabilistic framework adopted here substantially increases the sample size without introducing systematic disagreement in the overlapping subset. This confirms that the method provides a robust and statistically well-motivated extension of previous analyses. Beyond hard probability cuts, $P_{\rm ej}$ can also be used directly as a statistical weight in population or cross-correlation analyses, allowing uncertain events to contribute according to their model-dependent consistency with an origin in the effective core region.

        \subsection{Interpretation of the Core Size Parameter and Ejection Probability} \label{s:discussion_interp}
            
            For given trajectory models and effective core-region calibration, $P_{\rm ej}$ gives the probability that the closest-approach distance of a component trajectory is smaller than the adopted effective core radius. This probability is conditional on the trajectory model, its uncertainties, and the assumed core-region model.

            The core scale parameter $\sigma_{\rm c}$ should therefore not be interpreted exactly as the size of a specific physical region. Instead, it represents an effective observational scale that can schematically be written as
            \begin{equation}
            \sigma_{\rm c}^{2}
            \simeq
            \sigma_{\rm size}^{2}
            +
            \sigma_{\rm wander}^{2}
            +
            \sigma_{\rm other}^{2},
            \end{equation}
            where $\sigma_{\rm size}$ includes the finite VLBI-core size and its variability, $\sigma_{\rm wander}$ describes core-position variability, to which our method is primarily sensitive perpendicular to the local jet direction, and $\sigma_{\rm other}$ includes remaining effects such as trajectory-model uncertainties and blending.
            
            We estimate the core-size contribution using the fitted 15~GHz core sizes from \citet{Homan2021} for sources with estimated ejection epochs. The cores are modelled as elliptical Gaussians, with the full widths at half maximum reported for the minor and major axes. We convert these values to Gaussian standard deviations and use the minor-axis standard deviation as the core-size estimate, since it is less affected by unresolved extended jet emission. Epochs with only upper limits on the core size are excluded. Using the per-source mean core size and its standard deviation gives an effective population-level contribution of $\sigma_{\rm size}\simeq0.033$~mas. The estimate is based on approximately 10\% of all epochs with available core-size measurements, with a median coverage of approximately 6\% per source, providing a representative sampling of the population.

            Since all remaining contributions to $\sigma_{\rm c}$ are non-negative in variance, the measured $\sigma_{\rm c}=0.16\pm0.02$~mas also provides an upper limit on the core wander. Assigning the entire remaining variance to core-position variability gives
            \begin{equation}
            \sigma_{\rm wander}
            \lesssim
            \sqrt{\sigma_{\rm c}^{2}-\sigma_{\rm size}^{2}}
            \simeq 0.16~{\rm mas}.
            \end{equation}
            This value should therefore be regarded as a population-level upper limit on the characteristic core wander perpendicular to the jet direction. For comparison, \citet{Lisakov2017} measured a total displacement of $\simeq0.17$~mas of the 43~GHz core in 3C~273 during a strong flare, while \citet{Plavin2019}, using multi-epoch 2 and 8~GHz observations of 40 AGNs, found typical variations of individual core positions of about $0.35$~mas and a typical core-shift variability amplitude of about $0.4$~mas. These estimates probe somewhat different statistical quantities: the former is a direct displacement in a single source, while the latter is based on time-resolved core-shift variability in a sample of 40 sources. Our constraint, $\sigma_{\rm wander}\lesssim0.16$~mas, represents an upper limit on the characteristic core-position scatter perpendicular to the local jet direction inferred from jet-component trajectories of 394 sources. The previous measurements are therefore consistent with our wider population-level constraint, despite the different observing frequencies and statistical definitions. To our knowledge, this is the first such constraint obtained from the distribution of VLBI component trajectories over a large AGN sample. 

            Such a constraint may also be relevant for VLBI geodesy and astrometry, where compact AGNs are used as fiducial reference sources. A core shift \citep[e.g.,][]{Pushkarev2012} following exactly $r_{\rm core}\propto\nu^{-1}$ does not bias group-delay positions \citep{Porcas2009}, whereas time-variable departures from this relation can affect astrometric positions \citep{Plavin2019,Xu2025}. Our single-frequency constraint does not probe this frequency-dependent behaviour directly, but limits the characteristic 15~GHz core-position variability perpendicular to the jet.

            The effective scale may additionally depend on source geometry and redshift, since a fixed angular scale corresponds to different physical scales and 15~GHz observations probe different rest-frame frequencies. We nevertheless adopt a single $\sigma_{\rm c}$ for the full MOJAVE sample as a population-level simplification. Within this common calibration, $P_{\rm ej}$ can be used directly as an ejection probability, or as a model-dependent reliability weight when the absolute calibration of $\sigma_{\rm c}$ is uncertain.

            The method remains conditional on the adopted kinematic and core-region models. Incorrect component cross-identification or unresolved blending can bias the inferred closest-approach parameters. Additional biases may arise from neglected trajectory curvature, core-position variability, or underestimated trajectory uncertainties. These effects can in turn affect $P_{\rm ej}$. The population-level value of $\sigma_c$ derived here is specific to the MOJAVE 15~GHz sample. It should therefore be recalibrated for other frequencies, angular resolutions, or component-tracking procedures.

\section{Summary}
    \label{s:summary}

    We present a probabilistic framework for estimating ejection epochs of VLBI jet components from fitted trajectories. The ejection epoch is defined through the closest approach of the extrapolated trajectory to the VLBI core, while uncertainties in the trajectory parameters are propagated with Monte Carlo sampling. For each component, the method assigns an ejection probability $P_{\rm ej}$ from the probability that its closest-approach position lies within an adopted effective core region.
    
    Separately, using the distribution of closest-approach positions in the MOJAVE 15~GHz kinematic sample, we estimate the population-level scale of this region as $\sigma_{\rm c}=0.16\pm0.02$~mas, where $\sigma_{\rm c}$ is the standard deviation of the adopted circular Gaussian core-region model. In our model this corresponds to the average core radius $\bar{R}_{\rm c} = 0.20\pm0.03$~mas. From the measured VLBI core sizes, we estimate their contribution to be $\sigma_{\rm size}\simeq0.033$~mas, giving a population-level upper limit on the characteristic core wander perpendicular to the jet of $\sigma_{\rm wander}\lesssim0.16$~mas.

    Applying the method with this calibration to 1923 MOJAVE trajectories yields 1589 ejection epochs satisfying $t_{\rm ej}<t_{\rm first}$. Of these, 457 have $P_{\rm ej}\ge0.68$, compared with 284 ejection epochs reported by \citet{Lister2021}. The corresponding probability-weighted effective numbers are 387 and 222, respectively, giving a 74\% increase, while use of the full probabilistic sample yields $N_{\rm eff}=624$, a factor of 2.8 higher than the legacy value. For the overlapping subset, the resulting ejection epochs remain consistent with the earlier conservative estimates.
    
    The method remains conditional on the adopted trajectory and core-region models, and the use of a single angular $\sigma_{\rm c}$ for all sources is a population-level simplification. Nevertheless, it provides a homogeneous way to include uncertain ejection events and can be applied to other VLBI kinematic samples after recalibration of the effective core region.
        
\begin{acknowledgements}

    We thank the MOJAVE team for helpful discussions, including Alexander Plavin, Tigran Arshakian, Eduardo Ros, and Teresa Toscano Domingo for detailed comments on the manuscript.
    This research was funded by the European Union (ERC MuSES project No 101142396). Views and opinions expressed are however those of the author(s) only and do not necessarily reflect those of the European Union or the European Research Council. Neither the European Union nor the granting authority can be held responsible for them.
    This research has made use of data from the MOJAVE database that is maintained by the MOJAVE team \citep{Lister2018}.

\end{acknowledgements}

%


\bibliographystyle{bibtex/aa.bst}
\bibliography{references}{}






\end{document}